\documentclass[12pt,a4paper]{article}
\usepackage{amssymb,amsfonts,amsmath,amsthm}
\usepackage{hyperref}
\newtheorem{statement}{Statement}
\newtheorem{theorem}{Theorem}
\newtheorem{definition}{Definition}

\renewcommand{\AA}{\mathrm{A}}
\newcommand{\FF}{\mathrm{F}}
\newcommand{\dd}{\mathrm{d}}
\newcommand{\e}{\mathrm{e}}
\newcommand{\ra}{\,\rightarrow\,}
\newcommand{\g}{\mathfrak{g}}
\renewcommand{\phi}{\varphi}
\newcommand{\C}{\mathbf{C}}
\newcommand{\F}{\mathbf{F}}
\newcommand{\dt}{\delta}
\newcommand{\Z}{\mathbf{Z}}
\newcommand{\<}{\langle}
\renewcommand{\>}{\rangle}
\newcommand{\G}{\mathbf{G}}
\newcommand{\pt}{\partial}
\newcommand{\la}{\lambda}
\newcommand{\B}{\mathbf{B}}
\newcommand{\A}{\mathbf{A}}
\renewcommand{\t}{\mathfrak{t}}
\newcommand{\Ind}{\mathrm{Ind}\,}
\newcommand{\tg}{\widetilde{\g}}
\newcommand{\tG}{\widetilde{G}}
\newcommand{\teta}{\widetilde{\eta}}
\renewcommand{\H}{\mathbf{H}}
\newcommand{\txi}{\widetilde{\xi}}
\newcommand{\f}{\mathbf{f}}
\newcommand{\Ad}{\mathrm{Ad}}
\newcommand{\eps}{\epsilon}
\newcommand{\tH}{\widetilde{H}}
\newcommand{\ee}[2]{e^{#1}\wedge e^{#2}}
\renewcommand{\d}{\sigma}
\renewcommand{\k}{\kappa}

\author{Alexey A. Magazev, Igor V. Shirokov,  Yuriy Y. Yurevich}
\title{Integrable magnetic geodesic flows on Lie groups}
\date{}

\begin{document}
\maketitle

\begin{abstract}
Right-invariant geodesic flows on manifolds of Lie groups associated with 2-cocycles of corresponding Lie algebras are discussed. Algebra of integrals of motion for magnetic geodesic flows is considered and necessary and sufficient condition of integrability in quadratures is formulated. Canonic forms for 2-cocycles of all 4-dimensional Lie algebras are given and integrable cases among them are separated.
\end{abstract}

\section{Introduction}

The integration of geodesic flows on Riemannian and pseudo-Riemannian manifolds is a branch of modern mathematics and mechanics with a long history. Although many important and fundamental facts in this field were obtained in the classic works of the last century, this direction is still topical and is being developed currently. Indeed, the increased interest in this field by experts is indicated by the relatively large number of papers appearing recently (see, e.g., \cite{BolJov03} and the references therein). New examples of integrable geodesic flows and the classes of manifolds or metrics admitting such flows are especially interesting.

In addition to classical geodesic flows describing the motion of a free particle in a Riemannian (pseudo-Riemannian) space, there is an increased interest in the so-called \textit{magnetic geodesic flows}, which can be treated as the equations of motion of a particle on a Riemannian (pseudo-Riemannian) manifold in the presence of an external magnetic (electromagnetic) field. In this connection, we note two main approaches
used to describe geodesic motion in an external field. In the first approach, which is mainly adopted in theoretical physics, the electromagnetic field is described by a \textit{vector potential}, which is a 1-form $\AA$ on the manifold such that its exterior differential is the Faraday tensor of the electromagnetic field, $\FF = \dd \AA$. In this case, the equations of geodesic motion are modified by introducing new momenta $P_i$ related to the old momenta as $P_i = p_i + \e\, \AA_i$. In the second approach, which is more geometric and hence certainly more elegant, the electromagnetic field on a manifold $M$ is "turned on"\ via a deformation of the canonical Poisson bracket or, equivalently, via a deformation of the natural symplectic form $\omega$ of the cotangent bundle $T^* M$, $\omega \ra \omega +\e\, \pi_* \FF$, where $\FF$ is a closed 2-form on $M$ and $\pi$ is the natural projection $T^* M \ra M$  \cite{DubKriNov85}. Obviously, the second approach is more general, although the Poincare lemma implies that the two approaches are equivalent locally.

In this paper, we consider the problem of integrating magnetic geodesic flows on Lie group manifolds with right-invariant pseudo-Riemannian metrics. Although we investigate a special case of homogeneous spaces (Lie groups), the class of integrable magnetic geodesic flows on them turns out to be quite wide. In this paper, we consider \textit{arbitrary} closed right-invariant 2-forms on Lie groups associated with the corresponding spaces of 2-\textit{cocycles} of Lie algebras. A particular case of the proposed theory is given by magnetic geodesic flows associated with trivial 2-cocycles, whose restrictions to the coadjoint orbits are given by the well-known Kirillov forms. Up to now, most works on this subject were focused on these cases (see, e.g., \cite{Efi04}, \cite{Efi05}, \cite{BolJov06}), but nontrivial cocycles are in fact much more interesting from the standpoint of integration. In particular, as we show in what follows, for an arbitrary algebra and some 2-cocycle on it, there exists a so-called \textit{cohomology} index that generalizes the notion of the standard Lie algebra index and allows formulating a criterion for the integrability of magnetic geodesic flows in quadratures. Based on this criterion, we select all integrable cases among four-dimensional Lie groups and also give the canonical forms of the corresponding 2-cocycles.

\section{Magnetic geodesic flow on Lie groups with right-in\-va\-ri\-ant metrics}

Let $G$ be a connected real Lie group with the Lie algebra $\g$. We let $\rho_L$ denote the left regular representation of $\g$ acting on functions in $C^{\infty}(G)$. Because generators of the left regular representation of $G$ are right-invariant vector fields, the action of an operator in the representation $\rho_L$ can be written as
\begin{equation*}
\rho_L(X)\, \varphi(g) = \eta_X (\phi),\quad \phi \in C^{\infty}(G),
\end{equation*}
where $\eta_X \equiv -(R_g )_*\, X$ is the right-invariant vector field in the direction of $X \in \g$. 

We let $\C^2 (\g; C^{\infty}(G))$ denote the linear space of 2-\textit{cochains} over the module $C^{\infty}(G)$, i.e., the space of bilinear skew-symmetric functions on the Lie algebra $\g$ with values in $C^{\infty}(G)$. An arbitrary 2-cochain $\F \in \C^2(\g; C^{\infty}(G))$ bijectively corresponds to a dfferential 2-form $\FF \in \Omega^2(G)$ on $G$ acting on arbitrary tangent vectors 	$\zeta_1, \zeta_2 \in T_g G$ in accordance with the formula
\begin{equation}
\label{eq:01}
\FF(\zeta_1,\zeta_2)=\F((R_{g^{-1}})_*\, \zeta_1,(R_{g^{-1}})_*\, \zeta_2).
\end{equation}
We consider a closed form $\FF$ such that $\dd \FF = 0$. In this case, the corresponding 2-cochain $\F$ satisfies the condition
\begin{equation}
\label{eq:02}
\sum \limits_{XYZ \, \circlearrowleft} \eta_X\,\F(Y,Z)+\F(X,[Y,Z])=0,
\end{equation}
where the summation ranges all cyclic permutations of arbitrary vectors $X, Y, Z \in \g$. The obtained condition can be written more compactly in terms of the \textit{coboundary operator} $\dt: \C^k(\g; C^{\infty}(G)) \ra \C^{k+1}(\g; C^{\infty}(G))$ such that $\dt^2 = 0$ \cite{GotGro78}. Indeed, it can be shown that equality (\ref{eq:02}) is equivalent to the condition $\dt \F = 0$, whereby the 2-cochain $\F$ is a so-called 2-\textit{cocycle}. The set of all 2-cocycles over the module $C^{\infty}(G)$ is a subspace in $\C^2(\g; C^{\infty}(G))$ and is denoted by $\Z^2(\g; C^{\infty}(G))$. We consider only closed 2-forms in what follows.

We fix a basis $\{e_a \}$ in the Lie algebra $\g$ and the corresponding dual basis $\{e^a \}$ in the dual space $\g^*$: $\<e^a, e_b \> = \dt^a_b$, $a, b = 1, \dots , \dim \g$. In the chosen basis, the 2-cocycle $\F$ is written as
\begin{equation*}
\F=\frac12 \, \F_{ab}(g)\, e^a \wedge e^b,\quad \F_{ab}(g) \equiv \F(e_a,e_b),\quad g \in G,
\end{equation*}
and cocycle condition (\ref{eq:02}) as
\begin{equation}
\label{eq:03}
\sum \limits_{a,b,c \, \circlearrowleft} \eta_a\, \F_{bc} + C^d_{ab}\, \F_{cd}=0,
\end{equation}
where $\eta_a = -(R_g)_*\, e_a$ are basis right-invariant vector fields on $G$ and the $C_{ab}^c = -C_{ba}^c$ are the structure constants
of the Lie algebra in the chosen basis. Let $\{g^i\}$ be the local coordinates of an element $g \in G$. Then the components of the 2-form $\FF$ corresponding to the 2-cocycle $\F$ are given by
\begin{equation}
\label{eq:04}
\FF_{ij}(g) = \F_{ab}(g)\,\sigma^a_i(g)\, \sigma^b_j(g),
\end{equation}
where $\sigma_i^a (g)$ are components of the basis right-invariant 1-form $\sigma^a = -(R_g)^* e^a$.

On the algebra $\g$, we consider a nondegenerate quadratic form $\G$ that defines a pseudo-Riemannian metric on the group $G$ at unity. Acting with right shifts on the form $\G$, we obtain the right-invariant metric on the group at an arbitrary point $g \in G$:
\begin{equation}
\label{eq:05}
\mathrm{g}(\zeta_1,\zeta_2) = \G( (R_{g^{-1}})_*\, \zeta_1, (R_{g^{-1}})_*\, \zeta_2),\quad \zeta_1,\zeta_2 \in T_g \,G.
\end{equation}
The components of the right-invariant metric in terms of coordinates can be easily written in terms of the basis right-invariant 1-forms:
\begin{equation*}
\mathrm{g}_{ij}(g)= \G_{ab}\, \sigma^a_i(g)\, \sigma^b_j(g),
\end{equation*}
where $\G_{ab} \equiv \G(e_a , e_b)$ are the components of the form $\G$ in the chosen basis. In what follows, we choose the metric on the Lie group manifold $G$ as the right-invariant metric of form (\ref{eq:05}) (this metric is pseudo-Riemannian in general).

On the cotangent bundle $T^* G$, there exists the natural symplectic structure
\begin{equation}
\label{eq:06}
\omega = \dd p_i \wedge \dd g^i,
\end{equation}
which endows $T^* G$ with the structure of a symplectic manifold. We introduce a deformation of the symplectic form $\omega$ corresponding to turning on the external electromagnetic field described by closed 2-form (\ref{eq:04}):
\begin{equation}
\label{eq:07}
 \omega \rightarrow \omega_{\e} = \omega + \e\,\pi^* \FF.
\end{equation}
Here, $\pi$ is the natural projection $T^* G \ra G$, and $\e \in \mathbb{R}$ is a real parameter (charge). The form $\omega_{\e}$ is closed
by virtue of the condition $\dd \FF = 0$, and it therefore defines a new symplectic structure in $T^* G$. We assign an arbitrary function $\phi \in C^{\infty}(T^* G)$ the vector field $\nabla \phi$ such that
\begin{equation*}
\dd \phi (\zeta) = \omega_{\e} (\zeta, \nabla \phi),\quad \zeta \in T_{(g,p)}(T^* G).
\end{equation*}
In the space of functions $C^{\infty}(T^* G)$, we consider the Poisson bracket determined by the structure $\omega_{\e}$ as
\begin{equation}
\label{eq:08}
\{\phi, \psi \} = - \omega_{\e}(\nabla \phi, \nabla \psi).
\end{equation}
In local coordinates, the basis brackets corresponding to (\ref{eq:08}) are
\begin{equation}
\label{eq:09}
\{ g^i, g^j \} = 0, \quad  \{ p_i, g^j \} = \delta_i^j, \quad \{ p_i, p_j \} = \e\,\FF_{ij}(g).
\end{equation}
As is easily seen, these are deformations of the corresponding basis brackets for the standard canonical Poisson bracket corresponding to the standard form $\omega$.

We consider the function (a Hamiltonian)
\begin{equation}
\label{eq:10}
H(g,p) = \frac12\, |p|^{\,2} = \frac12\, \mathrm{g}^{ij}(g)\, p_i\, p_j,\quad p \in T^*G,
\end{equation}
on the cotangent bundle $T^* G$. The Hamiltonian system with Hamiltonian (\ref{eq:10}) defined on the symplectic manifold $(T^*G, \omega_{\e})$ is called the \textit{magnetic geodesic flow} on the Lie group associated with the form $\FF$. It is written in local coordinates as
\begin{equation}
\label{eq:11}
\dot{g}^i= \frac{\pt H(g, p)}{\pt p_i},\quad \dot{p_i} = - \frac{\pt H(g, p)}{\pt g^i} - \e \, \FF_{ij}(g)\, \frac{\pt H(g, p)}{\pt p_j}.
\end{equation}

\section{Vector potential of the right-invariant electromagnetic field tensor}

In the general case, the deformation of a symplectic structure corresponding to turning on an external electromagnetic field can essentially change the algebra of the integrals of motion of the original geodesic flow. Clearly, analyzing the general situation is a rather complicated problem because the integrability of the original flow is lost in the majority of cases. We therefore restrict ourself here to the situation where turning on the field preserves the number of the original symmetries in the problem.

The action of a Lie group $G$ on itself by right shifts induces the corresponding group action on the cotangent bundle $G: T^* G \ra T^* G$; this action preserves natural symplectic structure (\ref{eq:06}). But in terms of the form $\omega_{\e}$, which is the result of deforming the standard form $\omega$, the action of $G$ is no longer symplectic in general. We consider a right-invariant 2-form $\FF$, i.e., $(R_g)^*\, \FF = \FF$. It is easy to show that the corresponding 2-cocycle must then be invariant under the right regular representation,
\begin{equation}
\label{eq:12}
T^R_g \F(\cdot,\cdot) = \F(\cdot,\cdot), \quad g \in G,
\end{equation}
and must therefore be entirely determined by its value at the unity of the group. The set of all 2-cocycles satisfying condition (\ref{eq:12}) constitutes a subspace in $\Z^2(\g; C^{\infty}(G))$ isomorphic to the space $\Z^2(\g; \mathbb{R})$ of all 2-cocycles over the trivial module $\mathbb{R}$. In this case, the components of the 2-cocycle $\F$ are constant functions, which allows considerably simplifying cocycle condition (\ref{eq:03}):
\begin{equation}
\label{eq:13}
\sum \limits_{abc \, \circlearrowleft}\, C^d_{ab} \F_{cd}=0.
\end{equation}

Equation (\ref{eq:13}) is a homogeneous linear algebraic equation for the components of the 2-cocycle $\F$. A solution of this equation is the so-called \textit{trivial} 2-cocycle, given by $\F_{\la} = \dt \la$, where $\la \in \g^*$. Such 2-cocycles span the subspace $\B^2(\g; \mathbb{R}) \subset \Z^2(\g; \mathbb{R})$ of 2-\textit{coboundaries} of $\g$. Trivial 2-cocycles are the best studied examples of forms describing an electromagnetic field on Riemannian $G$-spaces \cite{Efi05}, \cite{BolJov06} and are a rather trivial case from the standpoint of integrability. Indeed, as we see in what follows, the algebra of the integrals of motion of the magnetic geodesic flow corresponding to a trivial 2-cocycle is actually isomorphic to the algebra of the integrals $\g$ of the original geodesic flow in the absence of the field. 

We address the following question. What is the vector potential $\AA$ corresponding to the electromagnetic field tensor $\FF = \dd \AA$? In the language of cochains, answering this question amounts to finding a 1-cochain $\A \in \C^1(\g; C^{\infty})$ such that $\dt \A = \F$ or, in more detail,
\begin{equation}
\label{eq:14}
\eta_X \A(Y) - \eta_Y \A(X) - \A([X,Y]) = \F(X,Y), \qquad X,Y \in \g.
\end{equation}

Let $\F = \dt \la$ be a trivial 2-cocycle. It is obvious that in this case, the general solution of Eq. (\ref{eq:14}) is given by $\A = -\la + \dt \phi$, where $\phi$ is an arbitrary function belonging to $C^{\infty}(G)$ (a 0-cochain), and $\dt \phi (X) \equiv \eta_X (\phi)$. In what follows, we assume that the 2-cocycle $\F$ is nontrivial in general.

We consider the space $\tg = \g \oplus \t$, where $\dim \t = 1$ and $\t$ is the one-dimensional center of  $\tg$. Let $Z$ be a nonvanishing
element from $\t$. The commutation operation in the algebra $\tg$ can be written using the cocycle $\F$ as
\begin{equation}
\label{eq:15}
[X, Y]_{\F} = [X,Y] + \F(X,Y) Z, \qquad X,Y \in \g.
\end{equation}
Let $\tG$ be the connected Lie group with the Lie algebra $\tg$; $\tG$ is the central extension of $G$ with a closed one-dimensional center $T$. We consider the functional space
\begin{equation}
\label{eq:16}
\Ind_{\mathbb{R}}^{\infty}(\tG) = \{ \phi \in C^{\infty}(\tG) \,|\, \phi(t^{-1} g) = U(t)\, \phi(t),\ t \in T,\, g \in \tG \},
\end{equation}
where $U(t)$ is a one-dimensional continuous representation of the group $T$ in $\mathbb{R}$. Definition (\ref{eq:16}) can also be written in the infinitesimal form
\begin{equation}
\label{eq:17}
\Ind_{\mathbb{R}}^{\infty}(\tG) = \{ \phi \in C^{\infty}(\tG) \,|\,  \teta_Z\, \phi = U_*(Z)\, \phi \}
\end{equation}
where $\teta_Z$ denotes the right-invariant vector field on $\tG$ in the direction of the vector $Z \in \t$. We note that $\Ind^{\infty}_{\mathbb{R}} (\tG)$ can also be interpreted as the space of smooth sections of the vector $\tG$-bundle with the base $G$ and fiber $\mathbb{R}$.

\begin{statement}
\label{stat:1}
The general solution of Eq. (\ref{eq:14}) is given by
\begin{equation}
\label{eq:18}
\A(X) = U^{-1} \teta_X U \big |_G + \dt \phi(X),
\end{equation}
where $\teta_X$ is the right-invariant vector field on the group $\tG$ in the direction of $X \in \g$ and $\phi$ is an arbitrary function belonging to $C^{\infty}(G)$.
\end{statement}

\begin{proof}
Because $\dt \phi$ is the general solution of the corresponding homogeneous equation $\dt \A = 0$, we must in fact show that the function $U^{-1} \teta_X U \big |_G$ is a particular solution of Eq. (\ref{eq:14}). We set $\A'(X) \equiv U^{-1} \teta_X U \big |_G$ and first verify that this function is well-defined on $G$. Indeed, because the space $\Ind_{\mathbb{R}}^{\infty}(\tG)$ is invariant under the left regular representation of $\tG$, the action of an arbitrary right-invariant vector field $\teta_X$ on a function $U(t)$, where $X \in \g$, agrees with definition (\ref{eq:16}), and hence 
\begin{equation}
\label{eq:19}
\teta_X U = U \left( U^{-1} \teta_X U \right),
\end{equation}
where $U^{-1} \teta_X U$ is a function independent of the point in the fiber $T$, i.e., a function constant on each coset $G \simeq \tG/T$.

We next consider the following operator identity reflecting the commutation relations of the Lie algebra $\g^R(\tG)$ of right-invariant vector fields on the group $\tG$:
\begin{equation}
\label{eq:20}
\teta_X \teta_Y - \teta_Y \teta_X = \teta_{[X,Y]} + \F(X, Y)\, \teta_Z, \quad X,Y \in \g, \ Z \in \t.
\end{equation}
We act on the function $U(t)$, $t \in T$, with each side of this identity. Using (\ref{eq:19}) and multiplying by $U^{-1}$ from the left, we then obtain
\begin{equation}
\label{eq:21}
\teta_X \left( U^{-1} \teta_Y U \right ) - \teta_Y \left( U^{-1} \teta_X U \right ) = U^{-1} \teta_{[X,Y]} U + \F(X,Y).
\end{equation}
Projecting (\ref{eq:21}) on $G \simeq \tG/T$ and noting that $\teta_X \big |_G= \eta_X$, we finally obtain the equality
\begin{equation*}
\eta_X \A'(Y) - \eta_Y \A'(X) = \A'([X,Y]) + \F(X,Y),
\end{equation*}
where $\A'(X) \equiv U^{-1} \eta_X U \big |_G$; this equality holds identically for any $X, Y \in \g$.
\end{proof}

In the formulation of the statement just proved, we gave an explicit construction of a 1-cocycle $\A$ whose associated diﬀerential form $\AA$ is the vector potential of right-invariant Faraday tensor (\ref{eq:04}). But we note the following. Because we do not consider global issues, such as those related to topological characteristics of the group manifold $G$, in this paper, formula (\ref{eq:18}) is only a local construction. Studying global properties of the 1-form $\AA$ requires a separate investigation, amounting to the study of the second cohomology group $H^2(G) \simeq \H^2(\g; C^{\infty}(G))$.

To conclude this section, we give explicit formulas for the components of the cochain $\A$ and the corresponding vector potential $\AA$. For this, we represent an arbitrary element of $\tG$ as $\widetilde{g} = s(g) t$, where $t \in T$, $g \in G$ and $s$ is a smooth local section $s: G \ra \tG$ of the principal bundle $(\tG, G, \pi , T)$, $\pi \circ s = \mathrm{id}$. Let $\{g^0\}$ be local coordinates in the fiber $T$ and $\{g^i\}$ be the coordinates of an element $g$ of $G$. In these coordinates, right-invariant vector fields on $G$ have the form
\begin{equation}
\label{eq:22}
\teta_0(g) = - \frac{\pt}{\pt g^0},\qquad \teta_a(g) = \eta_a^i(g) \frac{\pt}{\pt g^i} + \teta^{\,0}_a(g) \frac{\pt}{\pt g^0}.
\end{equation}
Choosing a representation of $T$ as $U(t) = e^{- t}$ and using formulas (\ref{eq:18}) and (\ref{eq:22}), we finally obtain
\begin{equation}
\label{eq:23}
\A_a(g) = - \teta_a^{\,0}(g) + \eta_a\, \phi, \quad \AA_i(g) = - \teta_a^{\,0}(g)\,\sigma_i^a(g) + \frac{\pt \phi }{\pt g^i}.
\end{equation}

\section{The algebra of the integrals of motion of a magnetic geodesic flow}

The functions in $C^{\infty}(T^* G)$ that are invariant under Hamiltonian flow (\ref{eq:11}) are called \textit{integrals of motion} of the magnetic geodesic flow. Obviously, such is any function $I(g, p)$ that commutes with the Hamiltonian,
\begin{equation*}
\dot{I} = \{ I, H \} = 0.
\end{equation*}

We now investigate the algebra of the integrals of motion of the magnetic geodesic flow on a group $G$. It is known that the integrals of motion of the geodesic flow with an arbitrary right-invariant metric in the absence of a field are the left-invariant functions of the form
\begin{equation*}
\xi_X (g, p) \equiv p(\xi_X) = X^a \xi_a^i(g)\, p_i,
\end{equation*}
where $\xi_X =  (L_g )_* X$ is a left-invariant vector field on $G$ in the direction of $X \in \g$. Hence, the algebra of the integrals of motion of the geodesic flow with right-invariant metric (\ref{eq:05}) is isomorphic to the Lie algebra $\g$ of $G$. But the picture changes somewhat when the external field given by the right-invariant form (\ref{eq:01}) is turned on. It is easily seen that the left-invariant functions $\xi_X (g, p)$ in this case are no longer integrals of motions of magnetic flow (\ref{eq:11}). Indeed, a function $\xi(g, p) \in C^{\infty}(T^* G)$ is an integral of motion of the geodesic flow with an arbitrary right-invariant metric if and only if it commutes with an arbitrary right-invariant function $\eta_X (g, p) \equiv p(\eta_X)$. But
\begin{equation*}
\{\eta_X(g,p), \xi_Y(g,p)\} = \e \FF(\eta_X,\xi_Y),
\end{equation*}
which is nonzero in general. In other words, the group action on the symplectic manifold $T^* G$ with the form $\omega_e$ is not \textit{Hamiltonian}.

We consider the deformation of a left-invariant function a 1-cochain $\f \in \C^1(\g, C^{\infty}(G))$:
\begin{equation*}
\xi_X(g,p) \ra \xi^{(\e)}_X(g,p) \equiv \xi_X(g,p) + \e\,\f(X), \quad X \in \g.
\end{equation*}
It is easy to show that the necessary and sufficient condition for the functions $\xi^{(\e)}_X(g,p)$ to be integrals of motion of magnetic geodesic flow (\ref{eq:11}) is that the equality
\begin{equation}
\label{eq:24}
\eta_X \f(Y) + F(\eta_X, \xi_Y) = 0,
\end{equation}
be satisfied (this is a first-order diﬀerential equation for the function $\f(Y)$). We let $\txi_Y$ denote the left-invariant vector field on the central extension $\tG$ of $G$ in the direction of $Y \in \g$ and $U(t)$ denote the one-dimensional representation of $T$ in $\mathbb{R}$. We then have the following fact.

\begin{statement}
There exists a 1-cochain $\f \in  \C^1(\g, C^{\infty}(G))$ that is a solution of Eq. (\ref{eq:24}) and has the form
\begin{equation}
\label{eq:25}
\f(Y) = U^{-1} \left( \txi_Y + \teta_{\Ad_g Y} \right) U \big |_G. 
\end{equation}
\end{statement}

\begin{proof}
That the function $\f(Y )$ is well defined is easily proved similarly to how this was done in the proof of Statement \ref{stat:1}. It therefore remains to verify that $\f(Y)$ satisfies Eq. (\ref{eq:24}) for an arbitrary $Y \in \g$.

Because left- and right-invariant vector fields commute on any Lie group, we have the operator identity
\begin{equation}
\label{eq:26}
\teta_X \txi_Y - \txi_Y \teta_X=0,
\end{equation}
which holds for arbitrary $X,Y \in \g$. Using it, we can verify the equality
\begin{equation*}
\teta_X \left ( U^{-1} \txi_Y U \right ) = \txi_Y \left( U^{-1} \teta_X U \right ),
\end{equation*}
or, after projecting on $G$, the equality
\begin{equation}
\label{eq:27}
\eta_X \left ( U^{-1} \txi_Y U \big |_G \right ) = \left ( \xi_Y \A' \right )(X).
\end{equation}
As in the preceding section, we here use the notation $\A' (X) = U^{-1} \eta_X U \big |_G$.

We consider the action of the right-invariant vector field $\eta_X$ on function (\ref{eq:25}). With (\ref{eq:27}), we then have
\begin{multline*}
\eta_X \f (Y) = \eta_X  \left( U^{-1} \txi_Y U \big |_G  + \A' (\Ad_g Y) \right ) = \left( \xi_Y \A'  \right) (X) +\eta_X \left ( \A'( \Ad_g Y ) \right) = \\
= - \left( \eta_{\Ad_g Y} \A' \right)(X) + \left( \eta_X \A'  \right) (\Ad_g Y ) - \A'([X,\Ad_g Y] ) = \delta \A'(X, \Ad_g Y).
\end{multline*}
In this chain of relations, we use the fact that $\eta_X (\Ad_g Y ) = - [X, \Ad_g Y ]$, $g \in G$, and also $\eta_{\Ad_g X} = - \xi_X$. To finish
the proof, it remains to note that $\delta \A' = \F$, whence
\begin{equation*}
\eta_X \f(Y) = \delta \A'(X, \Ad_g Y) = \F(X, \Ad_g Y) = - \FF(\eta_X, \xi_Y), \quad  X,Y \in \g.
\end{equation*}
\end{proof}

It is useful to write explicit expressions for the components of 1-cochains $\f$ in local coordinates. For this, we note that in terms of the coordinates $g^0, \{g^i\}$, left-invariant vector fields on $\tG$ are written as
\begin{equation*}
\txi_0(g) = \frac{\pt}{\pt g^0}, \quad \txi_a(g) = \xi_a^i(g) \frac{\pt}{\pt g^i} + \txi_a^{\,0}(g) \frac{\pt}{\pt g^0},
\end{equation*}
whence using (\ref{eq:23}), we obtain a local formula for (\ref{eq:25}):
\begin{equation}
\label{eq:28}
\f_a(g) = - \txi_a^{\,0}(g) - \left( \Ad_{g}\right)_a^b \teta_b^{\,0}(g).
\end{equation}

We now derive the commutation relations for the algebra of the integrals of motion $\xi^{(\e)}_X(g,p)$ of the magnetic geodesic flow. For this, we write the commutator of two arbitrary integrals of motion:
\begin{equation}
\label{eq:29}
\{ \xi^{(\e)}_X(g,p), \xi^{(\e)}_Y(g,p) \} = \xi_{[X,Y]}(g,p) + \e \left( \FF(\xi_X, \xi_Y) + \xi_X \f(Y) - \xi_Y \f(X) \right).
\end{equation}
The Jacobi identity for the Poisson bracket implies that the right-hand side of (\ref{eq:29}) is again a function of the integrals of motion of the magnetic flow. Indeed, in the Lie algebra $\g^L(\tG)$, we write the commutation relations of left-invariant vector fields on $\tG$:
\begin{equation}
\label{eq:30}
[\txi_X, \txi_Y] = \txi_{[X,Y]} + \F(X,Y)\, \txi_0.
\end{equation}
Acting on a function $U(t)$ in the representation with both sides and making transformations similar to those used in the proof of Statement \ref{stat:1}, we obtain
\begin{equation}
\label{eq:31}
\xi_X  \left( U^{-1} \txi_Y U \big |_G \right) - \xi_Y  \left( U^{-1} \txi_X U \big |_G \right) = U^{-1} \txi_{[X,Y]} U \big |_G - \F(X,Y).
\end{equation}
Next, setting $\A'(X) = U^{-1} \eta_X U \big |_G$, we have the chain of relations
\begin{multline}
\label{eq:32}
\xi_X \A'(\Ad_g Y) - \xi_Y \A'(\Ad_g X) = \\ 
= - \eta_{\Ad_g X} \A'(\Ad_g Y) + \A'([\Ad_g X, \Ad_g Y]) + \eta_{\Ad_g Y} \A'(\Ad_g X) - \A'([\Ad_g Y, \Ad_g X]) = \\
= \F(\Ad_g X, \Ad_g Y) + \A'(\Ad_g [X, Y]).
\end{multline}
Using (\ref{eq:31}) and (\ref{eq:32}) and recalling that $\F(\xi_X , \xi_Y) = \F(\Ad_g X, \Ad_g Y)$, we can transform (\ref{eq:29}) to the final
form
\begin{equation}
\label{eq:33}
\{ \xi^{(\e)}_X(g,p), \xi^{(\e)}_Y(g,p) \} = \xi^{(\e)}_{[X,Y]}(g,p) - \e\, \F(X, Y).
\end{equation}
The obtained expression is just the sought commutation relations in the algebra of the integrals of motion of magnetic geodesic flow (\ref{eq:11}). Comparing (\ref{eq:33}) with relations (\ref{eq:30}), it is easy to see that this algebra is isomorphic to the Lie algebra $\tg$ of the group $\tG$.

As a special case, it is interesting to consider the situation where the 2-cocycle $\F$ is trivial, i.e., $\F = \delta \lambda$ with $\lambda \in \g^*$. Commutation relations (\ref{eq:33}) then become
\begin{equation}
\label{eq:34}
\{ \xi^{(\e)}_X(g,p), \xi^{(\e)}_Y(g,p) \} = \xi^{(\e)}_{[X,Y]}(g,p) - \e\,  \<\lambda, [X,Y] \>.
\end{equation}
We make the shift $\xi^{(\e)}_X \ra \txi^{(\e)}_X \equiv \xi^{(\e)}_X - \e\, \<\lambda, X\>$ by an additive constant, thus redefining the function $\xi^{(\e)}_X$, and
rewrite (\ref{eq:34}) in the new notation:
\begin{equation*}
\{ \txi^{(\e)}_X(g,p), \txi^{(\e)}_Y(g,p) \} = \txi^{(\e)}_{[X,Y]}(g,p).
\end{equation*}
In this case, the Lie algebra $\tg$ is said to be a \textit{split extension} of the Lie algebra $\g$, and the corresponding section $r: \g \ra \tg$ such that $r(X) = X - \e\, \<\lambda, X\>$ defines an algebra isomorphism $r(\tg) \simeq \g$ \cite{GotGro78}.

\section{Integrability of the magnetic geodesic flow in quadratures}

For a magnetic geodesic flow with a right-invariant metric and electromagnetic field tensor (\ref{eq:04}), we investigated the existence of the corresponding algebra of the integrals of motion that would be isomorphic to the algebra $\tg$ in the general case in the preceding section. In this section, we give a criterion for the integrability of such systems in quadratures. We note that the material in this section essentially relies on
the results in \cite{MagShi03}.

On the group $\tG$, we choose a right-invariant quadratic form $B$ acting on cotangent vectors $\tau_1, \tau_2 \in T_{g}^* \tG$, $g \in \tG$,
as
\begin{equation}
\label{eq:35}
B(\tau_1, \tau_2) = (\pi^* \G^{-1})( (R_{g^{-1}})^* \tau_1, (R_{g^{-1}})^* \tau_2).
\end{equation}
Using the natural symplectic structure on the cotangent bundle $T^* \tG$,
\begin{equation*}
\widetilde{\omega} = \dd p_i \wedge \dd g^i + \dd p_0 \wedge \dd g^0,
\end{equation*}
we associate the function $\tH(g, p) = B(p, p)/2$ with form (\ref{eq:35}). Because Eq. (\ref{eq:35}) is independent of the point
in the fiber $T \simeq \tG/G$, the variable $g^0$ is cyclic for the Hamiltonian system associated with the Hamiltonian $\tH(g, p)$. We fix the value of the corresponding integral of motion $p_0 = - \e$. This Hamiltonian system then naturally decomposes into the two subsystems
\begin{equation}
\label{eq:36}
\dot{g}^0 = \frac{\pt \tH(g,p)}{\pt p_0}, \qquad\qquad\quad p_0 = - \e,
\end{equation}
\begin{equation}
\label{eq:37}
\dot{g}^i = \frac{\pt \tH(g,p)}{\pt p_i}, \qquad \dot{p}_i = -\frac{\pt \tH(g,p)}{\pt g^i}.
\end{equation}
In subsystem (\ref{eq:37}), we change the variables as $p_i \ra p_i + \e\, \AA_i$, where $\AA_i$ is the vector potential of the right-invariant Faraday tensor $\FF_{ij}$. It can be easily shown that after this transformation, subsystem (\ref{eq:37}) takes the form of a magnetic geodesic flow on $G$, which is entirely equivalent to system (\ref{eq:11}). Hence, the magnetic geodesic flow equations on the group $G$ with respect to right-invariant metric (\ref{eq:05}) are a part of Hamiltonian system (\ref{eq:36}) and (\ref{eq:37}) on the central extension $\tG$. Therefore, the question regarding the integrability of geodesic flow (\ref{eq:11}) entirely reduces to the question of the integrability of Eqs. (\ref{eq:36}) and (\ref{eq:37}).

\begin{statement}
An arbitrary right-invariant magnetic geodesic flow (\ref{eq:11}) on a group $G$ is integrable in quadratures if and only if the right-invariant Hamiltonian system in Eq. (\ref{eq:37}) on the central extension $\tG$ of $G$ is integrable in quadratures.
\end{statement}

A constructive algorithm for integrating $G$-invariant Hamiltonian flows on homogeneous spaces in quadratures was proposed in \cite{MagShi03}; the corresponding necessary and sufficient conditions for the integrability were also derived there. An arbitrary Lie group $G$ can be considered a homogeneous space (for example, with respect to the right shifts $R_g : G \ra G$) with a trivial stationary group $H = \{e\}$. We therefore use
the results in \cite{MagShi03} and here formulate only the key points that allow stating the corresponding criterion for magnetic geodesic flows.

For an arbitrary group $G$, its Hamiltonian action on the cotangent bundle $T^* G$ with the natural symplectic structure $\omega = \dd p \wedge \dd x$ allows introducing the left moment map $M^L: T^* G \ra \g^*$, which is given by the relations
\begin{equation}
\label{eq:39}
M^L(X) = \xi_X(g,p), \quad M^L_a = X^a \xi_a(g,p), \quad X \in \g.
\end{equation}
The map in (\ref{eq:39}) is a Poisson map of the algebra $C^{\infty}(T^* G)$ to the algebra $C^{\infty}(\g^*)$ with the Poisson-Lie
bracket
\begin{equation}
\label{eq:40}
\{ \phi, \psi \}^{Lie} = \< f, [\dd \phi(f), \dd \psi(f) ] \>
\end{equation}
where $\phi, \psi \in C^{\infty}(\g^*)$ and $f \in \g^*$. Because bracket (\ref{eq:40}) is degenerate, the space $C^{\infty}(\g^*)$ may contain
\textit{Casimir functions} $K(f)$ that commute with any function on $\g^*$ , i.e., functions such that
\begin{equation*}
\{ K, \phi \}^{Lie} = 0 \qquad \text{for all } \phi \in C^{\infty}(\g^*).
\end{equation*}
We note that the number $\mathrm{ind}\, \g$ of independent Casimir functions is called the \textit{index} of $\g$ and is invariantly
defined as the dimension of the annihilator $\g^{\la}$ of a covector in the general position:
\begin{equation*}
\mathrm{ind}\, \g \equiv \min \limits_{\la \in \g^*} \dim \g^{\la}, \qquad \g^{\la} = \{ X \in \g \,|\, \<\la, [X, \g] \> = 0 \}.
\end{equation*}

The \textit{equivariance} property of the moment map in (\ref{eq:39}) allows assigning each Casimir function $K(f) \in C^{\infty}(\g^*)$ a function $K^L(g, p) \equiv K \circ M^L$ in $C^{\infty}(T^*G)$ that commutes with an arbitrary left-invariant function on $T^* G$. It is easy to see that all the functions $K^L(g,p)$ are in the center of the enveloping algebra $U(\g)$ and are integrals of motion of the right-invariant Hamiltonian flow on $G$. Similarly to map (\ref{eq:39}), we can introduce the right moment map $M^R: T^* G \ra \g^*$ as
\begin{equation}
\label{eq:41}
M^R(X) = \eta_X(g, p), \qquad M^R_a = X^a \eta_a^i(g) p_i, \quad X \in \g, \ M^R \in \g^*.
\end{equation}

We let $\Omega^L \subset \g^*$ and $\Omega^R \subset \g^*$ denote the symplectic leaves with respect to the left and right maps.
The Casimir functions $K^R (g, p) \equiv K \circ M^R$ are equal to the functions $K^L(g, p)$, and the centers of the Poisson algebras of left- and right-invariant functions in $C^{\infty}(T^* G)$ therefore coincide. This implies that $\Omega^R = \Omega^L \equiv \Omega$.

The Hamiltonian of a right-invariant flow on the group $G$ is a quadratic combination of right-invariant functions $H = H (\eta(g, p))$. It can be shown that the right moment map $M^R$ takes this flow into a Hamiltonian system with the Hamiltonian $H(M^R)$ with respect to bracket (\ref{eq:40}). Because Casimir functions exist in $C^{\infty}(\g^*)$, this Hamiltonian system can be restricted to the corresponding symplectic leaf $\Omega$, whose dimension is determined by the index of $\g$:
\begin{equation*}
\dim \Omega = \dim \g / \g^{\la} = \dim \g - \dim \g^{\la}.
\end{equation*}
Using this fact, we can formulate the following integrability criterion, which is a special case of a similar criterion obtained in \cite{MagShi03} and is applicable to an arbitrary homogeneous space.

\begin{statement}
\label{stat:3}
An arbitrary right-invariant Hamiltonian flow on a group $G$ reduces to a $\dim \Omega$-dimensional Hamiltonian system and in particular is integrable in quadratures if and only if
\begin{equation}
\frac{1}{2} \dim \Omega = \frac{1}{2} \left( \dim \g - \mathrm{ind}\, \g \right) < 2.
\end{equation}
\end{statement}

We now return to the investigation of the integrability of right-invariant magnetic geodesic flow (\ref{eq:11}). The assertion formulated in Statement \ref{stat:3} actually reduces the problem to finding an integrability condition for arbitrary right-invariant Hamiltonian systems on the central extension $\tG$ of $G$, which in turn leads to verifying the condition
\begin{equation}
\label{eq:43}
\frac{1}{2} \left( \dim \tg - \mathrm{ind}\, \tg \right) < 2.
\end{equation}
In this connection, we try to formulate this condition in terms of the Lie algebra $\g$ itself and the corresponding quotient cocycle $\F \in \Z^2(\g; \mathbb{R})$.

First, it is obvious that $\dim \tg = \dim \g + 1$. Next, in the dual space $\tg^* \simeq \g^* \oplus \t^*$ of the Lie algebra $\tg$, we fix 
a covector $\la \oplus \eps$ in the general position, where $\la \in \g^*$ and $\eps \in \t^*$ . Without loss of generality, we can assume that $\eps(e_0) = 1$, where $e_0$ is a basis vector of the one-dimensional algebra $\t$. Using structure (\ref{eq:15}) of commutation relations in $\tg$, it is easy to see that the one-dimensional center $\t$ entirely belongs to $\tg^{\la \oplus \eps}$, and this annihilator as a linear space can therefore be decomposed into the direct sum of subspaces $\g_{\F} \oplus \t$, where
\begin{equation}
\label{eq:44}
\g_{\F} \equiv \ker \F = \{ X \in \g \,|\, \F(X, \g) = 0 \}.
\end{equation}
In particular, it hence follows that $\dim \tg^{\la\oplus \eps} = \dim \g_{\F} + 1$. It is also easy to show that the subspace $\g_{\F}$ is
a subalgebra in $\g$ and that in the case where $\F = \dt \la \in \B^2(\g; \mathbb{R})$, it coincides with the standard annihilator $\g^{\la}$ of the covector $\la$.

\begin{definition}
Let $[\F] \in \H^2(\g; \mathbb{R})$ be a cohomology class of some 2-cocycle $\F \in \Z^2(\g; \mathbb{R})$. The index of $\g$ of the cohomology class $[\F]$ (or simply the cohomology index) is the number
\begin{equation}
\label{eq:45}
\mathrm{ind}_{[\F]}\, \g = \min \limits_{\F \in [\F]} \dim \g_{\F}.
\end{equation}
\end{definition}

This definition generalizes the notion of the standard index ($\mathrm{ind}\, \g$) of an algebra $\g$ and coincides with it in the particular case where the 2-cocycle $\F$ is trivial, i.e., $\F \in [0]$. Using this definition and inequality (\ref{eq:43}), it is now easy to formulate the integrability criterion in terms of the cohomology of the Lie algebra $\g$.

\begin{theorem}
On a group $G$, an arbitrary right-invariant magnetic geodesic flow (\ref{eq:11}) given by a 2-cocycle $\F \in \Z^2(\g; \mathbb{R})$ is integrable in quadratures if and only if
\begin{equation}
\label{eq:46}
\frac{1}{2}\, \left( \dim \g - \mathrm{ind}_{[\F]}\, \g \right) < 2.
\end{equation}
\end{theorem}

For semisimple Lie algebras, the Whitehead theorem holds, stating that $\H^2(\g; \mathbb{R}) = 0$ (see, e.g., \cite{GotGro78}).
Any 2-cocycle $\F$ of a semisimple $\g$ is trivial, whence $\mathrm{ind}_{[\F]}\, \g = \mathrm{ind}\, \g$. Therefore, the integrability property of geodesic flows for semisimple Lie groups is preserved under "turning on"\ external electromagnetic fields. Obviously, this statement remains valid in the case of an arbitrary Lie algebra $\g$ with a zero 2-cohomology group.

We give a nontrivial example illustrating the efficiency of the proposed integrability criterion. We consider the solvable four-dimensional Lie group $G$ whose algebra $\g$ has the nonvanishing commutation relations
\begin{equation*}
[e_1 , e_4] = e_4, \qquad  [e_2 , e_4 ] = e_4 .
\end{equation*}
It can be easily veriﬁed that $\mathrm{ind}\, \g = 2$, and an arbitrary right-invariant geodesic ﬂow on $G$ is therefore integrable in quadratures in the absence of an electromagnetic field.

We find the most general form of the 2-cocycle for this Lie algebra. For this, we solve the system of linear algebraic equations (\ref{eq:13}) to obtain the parametric family of solutions
\begin{equation}
\label{eq:47}
\F = \alpha\, e^1 \wedge e^2 + \beta\, e^1 \wedge e^3 + \gamma\, e^2 \wedge e^3 + \delta f,\quad \alpha, \beta, \gamma \in \mathbb{R}.
\end{equation}
Here, $f = f_4\, e^4$, whence $\delta f = f_4\, (e^1 + e^2 ) \wedge e^4$. It is easy to see that $\dim \H^2(\g; \mathbb{R}) = 3$ in our case.

We introduce local coordinates on $G$: $g = \exp(g^1 e_1)\exp(g^2 e_2)\exp(g^3 e_3)\exp(g^4 e_4)$. In these coordinates, we choose the basis of right-invariant 1-forms
\begin{equation*}
\sigma^1 = - \dd g^1,\quad \sigma^2 = - \dd g^2, \quad \sigma^3 = - \dd g^3, \quad \sigma^4 = - \exp(g^1 + g^2)\, \dd g^4 .
\end{equation*}
The electromagnetic field form $\FF$ corresponding to 2-cocycle (\ref{eq:47}) is given by
\begin{equation}
\label{eq:48}
\FF = \alpha\, \dd g^1 \wedge \dd g^2 + \beta\, \dd g^1 \wedge \dd g^3 + \gamma\, \dd g^2 \wedge \dd g^3 + f_4 \exp(g^1 + g^2)\, (\dd g^1 + \dd g^2) \wedge \dd g^4.
\end{equation}

We consider the central extension $\tg$ of $\g$ with the 2-cocycle $\F = \alpha\, e^1 \wedge e^2 + \beta\, e^1 \wedge e^3 + \gamma\, e^2 \wedge e^3$ in the general position. The nonvanishing commutation relations of this algebra can be written as
\begin{equation*}
[e_1, e_2] = \alpha\, e_0, \quad [e_1, e_3] = \beta\, e_0, \quad [e_1, e_4] = e_4, \quad [e_2, e_3] = \gamma\,e_0, \quad [e_2, e_4] = e_4,
\end{equation*}
where $e_0$ is the basis vector of the one-dimensional center $\t$. Let $g^0$ be the coordinate of an element in the center $T$ of $G$. In terms of the coordinates $\widetilde{g} = \exp(g^0 e_0) g$, $\widetilde{g} \in \tG$, $g \in G$, a basis of left- and right-invariant vector fields of the group $\tG$ can then be chosen as
\begin{equation*}
\txi_0 = \frac{\pt}{\pt g^0},\quad \txi_1 = \frac{\pt}{\pt g^1} - g^4 \frac{\pt}{\pt g^4} - (\alpha g^2 + \beta g^3) \frac{\pt}{\pt g^0},\quad \txi_2 = \frac{\pt}{\pt g^2} - g^4 \frac{\pt}{\pt g^4} - \gamma g^3 \frac{\pt}{\pt g^0},
\end{equation*}
\begin{equation*}
\txi_3 = \frac{\pt}{\pt g^3}, \quad \txi_4 = \frac{\pt}{\pt g^4},\quad \teta_0 = - \frac{\pt}{\pt g^0}, \quad \teta_1 = - \frac{\pt}{\pt g^1},\quad \teta_2 = - \frac{\pt}{\pt g^2} + \alpha g^1 \frac{\pt}{\pt g^0},
\end{equation*}
\begin{equation*}
\teta_3 = -\frac{\pt}{\pt g^3} + \left( \beta g^1 + \gamma g^2 \right) \frac{\pt}{\pt g^0},\quad \teta_4 = - \exp(- g^1 - g^2) \frac{\pt}{\pt g^4}.
\end{equation*}
We choose $U = \exp(- g^0)$ as the representation of the group $T$ in $\mathbb{R}$. Using formulas (\ref{eq:23}), we easily find that
\begin{equation*}
\A(g) = - \alpha g^1 e^2 - (\beta g^1 + \gamma g^2 )\, e^3,\quad \AA(g) = \alpha\, g^1\, \dd g^2 + (\beta g^1 + \gamma g^2 )\, \dd g^3.
\end{equation*}
By direct calculation, it is easy to verify that $\F = \delta \A$ and $\FF = \dd \AA$.

Similarly, using formula (\ref{eq:28}), we explicitly construct the algebra of the integrals of motion for an arbitrary right-invariant magnetic geodesic flow for this group:
\begin{equation*}
\xi_1^{(\e)}(g,p) = p_1 - g^4 p_4 + \e \left(\alpha g^2 + \beta g^3 \right),\quad \xi_2^{(\e)}(g,p) = p_2 - g^4 p_4 - \e \left( \alpha g^1 - \gamma g^3 \right),
\end{equation*}
\begin{equation*}
\xi_3^{(\e)}(g,p) = p_3 - \e \left(\beta g^1 + \gamma g^2\right), \quad \xi_4^{(\e)}(g,p) = p_4.
\end{equation*}
It is remarkable that for $\e = 0$, these functions are integrals of motion of the right-invariant geodesic flow on the Lie group in the absence of the field and constitute a subalgebra of left-invariant functions in $C^{\infty}(T^* G)$ with respect to the standard Poisson bracket.

In our example, we now select the cases of integrable magnetic geodesic flows. For this, recalling that $\g_{\F} = \ker \F$, we rewrite condition (\ref{eq:46}) in a somewhat different form:
\begin{equation*}
\frac{1}{2}\, \left( \dim \g - \mathrm{ind}_{[\F]}\, \g \right) = \frac{1}{2} \max \limits_{\F \in [\F]} \mathrm{rank}\, \F < 2.
\end{equation*}
Because the rank of a skew-symmetric bilinear form is an even number, we have
\begin{equation*}
\mathrm{rank}\,\F=\left\{%
\begin{array}{ll}
    4, & \hbox{$\beta \neq \gamma$;} \\
    2, & \hbox{$\beta=\gamma$.} \\
\end{array}%
\right.	
\end{equation*}
Therefore, the 2-cocycles admitting integrable magnetic geodesic flows have the form
\begin{equation*}
\F=\alpha\, e^1 \wedge e^2+\beta\, (e^1 + e^2) \wedge e^3+\delta f,\quad  f \in \g^*.
\end{equation*}
The corresponding Casimir functions on $\tg^*$ are given by
\begin{equation*}
K_0 = f_0, \quad K_1 = \beta (f_1 - f_2) + \alpha f_3, \quad K_2 = f_4^{\beta}\,e^{- f_3}.
\end{equation*}

To conclude, we emphasize once again that the magnetic geodesic flows on Lie groups considered here were of interest to us just from the standpoint of integrability in quadratures. It is well known that the term "integrability"\ is often used in a somewhat diﬀerent sense, that of the existence of suﬃciently many first integrals of motion satisfying special conditions (the \textit{complete integrability}). Naturally, the two notions are related, but these two terms are nevertheless not equivalent. We show this in a simple example. We consider a commutative Lie group whose manifold is the two-dimensional torus $\mathbf{T}^2 \simeq S^1 \times S^1$. We let $\phi$ and
$\psi$ denote angular coordinates on $\mathbf{T}^2$ parameterizing an arbitrary group element. For simplicity, we consider the ﬂat metric $\dd s^2 = \dd \phi^2 + \dd \psi^2$ on this Lie group and choose the 2-form proportional to the volume element as the electromagnetic field: $\FF = \e\, \dd \phi \wedge \dd \psi$. This form is closed but not exact on the entire manifold $\mathbf{T}^2$,
i.e., the corresponding vector potential $\AA$ does not exists globally. Moreover, it can be easily seen that the set of first integrals of motion of the magnetic geodesic flow are also deﬁned only locally and in the chosen coordinates have the form
\begin{equation*}
\xi_1^{(\e)}(g,p)=p_{\phi} - \e\, \psi,\quad \xi_2^{(\e)}(g,p) = p_{\psi} + \e\,\phi.
\end{equation*}
The magnetic geodesic flow in this example is therefore not completely integrable on the entire manifold. But it is easily integrated in quadratures; moreover, it is integrable globally:
\begin{equation*}
\phi(t) = p_{\phi}(0)\,\frac{\sin(\e\,t)}{\e} + p_{\psi}(0)\,\frac{\cos(\e\,t) - 1}{\e}+\phi(0),
\end{equation*}
\begin{equation*}
\psi(t) = - p_{\phi}(0)\,\frac{\cos(\e\,t) - 1}{\e} + p_{\psi}(0)\,\frac{\sin(\e\, t)}{\e} + \psi(0),
\end{equation*}
\begin{equation*}
p_{\phi}(t) = p_{\phi}(0) \cos(\e\,t) - p_{\psi}(0) \sin(\e\,t),
\end{equation*}
\begin{equation*}
p_{\psi}(t) = p_{\phi}(0) \sin(\e\,t) + p_{\psi}(0) \cos(\e\,t).
\end{equation*}

\section*{Appendix}

In this appendix, we give the canonical form of 2-cocycles $\F$ for all four-dimensional Lie algebras. The canonical forms are obtained by acting with the automorphism group of the corresponding algebras. In addition, we identify the integrable cases for each Lie algebra, i.e., the subclasses of 2-cocycles for which $\mathrm{ind}_{[\F]}\, \g = 2$. The chosen classiﬁcation of four-dimensional Lie algebras is given in Petrov’s book \cite{Pet61}. In what follows, we use the notation $\varepsilon, \sigma = \pm 1$, $\kappa = 0, \pm 1$, $\alpha, C \in \mathbb{R}$.

\begin{enumerate}
    \item Commutative algebra $\g_0$ : $[e_A,e_B]=0,\quad A,B=1,\dots,4$;\\
    $$
    \F=e^1 \wedge e^2 + \kappa\, e^3 \wedge e^4,\quad \mathrm{ind}_{[\F]}\,\g =2 \text{ при }
    \kappa=0.
    $$

\item
    Algebra $\g_1$: $[{e_{2}}, \,{e_{3}}]={e_{1}},$ $[{e_{1}}, \,{e_{4}}]=\alpha,$ ${e_{1}}, \,[{e_{2}}, \,{e_{4}}]={e_{2}},$ $[{e_{3}}, \,{e_{4}}]=( \alpha  - 1)\,{e_{3}}$;
$$
\F_1 = \d(\ee{1}{4} + \frac{1}{\alpha}\,\ee{2}{3}) + \k\,\ee{2}{4},\quad \alpha \neq 0;
$$
$$
\F_2 = \k_1\,\ee{2}{4} + \k_2\,\ee{3}{4},\quad \mathrm{ind}_{[\F_2]}\,\g=2\text{ for }
\alpha=0.
$$
\item
    Algebra $\g_2$:  $ [{e_{1}}, \,{e_{4}}]=2\,{e_{1}}, \,[{e_{3}}, \,{e_{4}}]={e_{2}}
 + {e_{3}}, \,[{e_{2}}, \,{e_{3}}]={e_{1}}, \,[{e_{2}}, \,{e_{4}}
]={e_{2}}$;
$$
\F_1 = \k(\ee{1}{4} + \frac12\,\ee{2}{3}),\quad \F_2 = \k\,\ee{2}{4},\quad \F_3 =
\k\ee{3}{4}.
$$
\item
    Algebra $\g_3$:  $ [{e_{2}}, \,{e_{4}}]={e_{3}}, \,[{e_{3}}, \,{e_{4}}]= - {e_{2}}
 + \alpha \,{e_{3}}, \,[{e_{2}}, \,{e_{3}}]={e_{1}}, \,[{e_{1}},
\,{e_{4}}]=\alpha \,{e_{1}}$;
$$
\F_1 = \d(\ee{1}{4} + \frac{1}{\alpha}\,\ee{2}{3}),\quad \alpha \neq 0;
$$
$$
\F_2 = \k\,\ee{2}{4} + C\,\ee{3}{4}.
$$
\item
    Algebra $\g_4$: $\quad [{e_{1}}, \,{e_{4}}]={e_{1}}, \,[{e_{2}}, \,{e_{3}}]={e_{2}}$;
    $$
    \F = \k_1\,\ee{1}{4} + \k_2\,\ee{2}{3} + C\,\ee{3}{4}.
    $$
\item
    Algebra $\g_5$: $\quad [{e_{2}}, \,{e_{4}}]= - {e_{1}}, \,[{e_{1}}, \,{e_{4}}]={e_{2}}
, \,[{e_{1}}, \,{e_{3}}]={e_{1}}, \,[{e_{2}}, \,{e_{3}}]={e_{2}}$;
$$
\F_1 = \d\,(\ee{1}{4} + \ee{2}{3}) + C\,\ee{3}{4};
$$
$$
\F_2 = \k\,(-\ee{1}{3} + \ee{2}{4}) + C\,\ee{3}{4}.
$$
\item
    Algebra $\g_6$: $\quad [{e_{1}}, \,{e_{4}}]={e_{1}}$;
$$
\F_1 = \k_1\ee{1}{4} + \d\ee{2}{3} + \k_2\ee{2}{4};
$$
$$
\F_2 = \k_1\ee{1}{4} + \k_2\ee{2}{4} + \k_3\ee{3}{4}.
$$
\item
    Algebra $\g_7$: $\quad [{e_{1}}, \,{e_{4}}]={e_{4}}, \,[{e_{2}}, \,{e_{4}}]={e_{4}}$;
$$
\F_1 = \k_1\ee{1}{3} + \k_2(\ee{1}{4} + \ee{2}{4}) + \k_3\ee{2}{3},\quad
\mathrm{ind}_{[\F_1]}\,\g =2 \text{ for } \k_2=\d;
$$
$$
\F_2 = \d\ee{1}{2} + \k_1(\ee{1}{4} + \ee{2}{4}) + \k_2\ee{2}{3},\quad
\mathrm{ind}_{[\F_2]}\,\g =2 \text{ for } \k_1=0;
$$
$$
\F_3 = \k_1(\ee{1}{4} + \ee{2}{4}) + \k_2\ee{2}{3}, \quad \mathrm{ind}_{[\F_3]}\,\g =2
\text{ for } \k_1=0.
$$
\item
    Algebra $\g_8$: $\quad [{e_{1}}, \,{e_{4}}]={e_{4}}, \,[{e_{2}}, \,{e_{4}}]={e_{4}},
\,[{e_{3}}, \,{e_{4}}]={e_{4}}$;
$$
\F_1 = \k_1\ee{1}{2} + \d\ee{1}{3} + \k_2(\ee{1}{4} + \ee{2}{4} + \ee{3}{4}),\quad
\mathrm{ind}_{[\F_1]}\,\g =2 \text{ for } \k_2=\d;
$$
$$
\F_2 = \k_1\ee{1}{2} + \k_2(\ee{1}{4} + \ee{2}{4} + \ee{3}{4}) + \k_3\ee{2}{3},\quad
\mathrm{ind}_{[\F_2]}\,\g =2 \text{ for } \k_2=\k_3.
$$
\item
    Algebra $\g_9$: $\quad [{e_{1}}, \,{e_{4}}]={e_{1}} + {e_{4}}$;
$$ \F_1 = \k_1(\ee{1}{3} + \ee{3}{4}) + \k_2\ee{1}{4} + \d\ee{2}{3}; $$
$$ \F_2 = \k_1(\ee{1}{2} + \ee{2}{4}) + \k_2(\ee{1}{3} + \ee{3}{4}) + \k_3\ee{1}{4},\quad \mathrm{ind}_{[\F_2]}\,\g =2.  $$
\item
    Algebra $\g_{10}$: $\quad [{e_{1}}, \,{e_{4}}]=\alpha \,{e_{1}} + {e_{2}}, \,[{e_{2}}, \,
{e_{4}}]=\alpha \,{e_{2}}, \,[{e_{3}}, \,{e_{4}}]={ e_{3}}$;
$$ \F_1 = \k_1\ee{2}{4} + \k_2\ee{3}{4},\quad  \mathrm{ind}_{[\F_1]}\,\g =2;$$
$$ \F_{2} = \k_1\ee{1}{4} + \k_2\ee{3}{4},\quad \mathrm{ind}_{[\F_2]}\,\g =2. $$
\item
    Algebra $\g_{11}$: $\quad [{e_{2}}, \,{e_{4}}]=\alpha \,{e_{2}} + {e_{3}}, \,[{e_{1}}, \,{e_{4}}]=\alpha
    \,{e_{1}} + {e_{2}}$;
$$ \F = \k\ee{1}{4} + C_1\ee{2}{4} + C_2\ee{3}{4},\quad  \mathrm{ind}_{[\F]}\,\g =2. $$
\item
    Algebra $\g_{12}$: $\quad [{e_{2}}, \,{e_{4}}]=\alpha \,{e_{2}} + {e_{3}}, \,[{e_{3}}, \,
{e_{4}}]={e_{3}}, \,[{e_{1}}, \,{e_{4}}]=\alpha \,{e_{1}} + {e_{2 }}$;
$$ \F = \k\ee{1}{4} + C_1\ee{2}{4} + C_2\ee{3}{4},\quad  \mathrm{ind}_{[\F]}\,\g =2. $$
\item
    Algebra $\g_{13}$: $\quad [{e_{2}}, \,{e_{4}}]= - {e_{1}} + \alpha \,{e_{2}}, \,[{e_{1}}
, \,{e_{4}}]=\alpha \,{e_{1}} + {e_{2}}, \,[{e_{3}}, \,{e_{4}}]= \varepsilon \,{e_{3}}$;
$$ \F{a_{13}}{} = \k_1\ee{1}{4} + C\ee{2}{4} + k_2\ee{3}{4},\quad  \mathrm{ind}_{[\F]}\,\g =2.  $$
\item
    Algebra $\g_{14}$: $\quad [{e_{1}}, \,{e_{2}}]={e_{1}}, \,[{e_{2}}, \,{e_{3}}]={e_{3}},
\,[{e_{1}}, \,{e_{3}}]=2\,{e_{2}}$;
$$ \F_{1} = C\ee{1}{2} + \k\ee{2}{3},\quad \mathrm{ind}_{[\F_1]}\,\g =2;  $$
$$ \F_{2} = C\ee{1}{3} + \k\ee{2}{3},\quad \mathrm{ind}_{[\F_2]}\,\g =2. $$
\item
    Algebra $\g_{15}$: $\quad [{e_{1}}, \,{e_{2}}]={e_{3}}, \,[{e_{1}}, \,{e_{3}}]= - {e_{2}}
, \,[{e_{2}}, \,{e_{3}}]={e_{1}}$;
$$ \F = C_1\ee{1}{2} + C_2\ee{1}{3} + C_3\ee{2}{3},\quad\mathrm{ind}_{[\F]}\,\g =2. $$
\end{enumerate}


\end{document}